\def\kms{km s$^{-1}$\ }

\documentstyle[12pt,emulateapj,amssym,flushrt,psfig]{article}

\begin{document}

\title{ Correlation Function of Galaxy Groups }
\vskip 1cm

\author{Manuel E. Merch\'an\altaffilmark{1},
 Marcio A. G. Maia\altaffilmark{2} and Diego G. Lambas\altaffilmark{1}}

\affil{manuel@oac.uncor.edu, maia@on.br, dgl@oac.uncor.edu}

\altaffiltext{1}{Grupo de Investigaci\'on en Astronom\'{\i}a Te\'orica y
Experimental (IATE), Observatorio Astron\'omico, Laprida 854, C\'ordoba and
CONICET, Argentina.}

\altaffiltext{2}{Departamento de Astromomia, Observat\'orio Nacional, Rua 
General Jos\'e Cristino 77, Rio de Janeiro, 20921-030, Brazil.}

\begin{abstract}

We use the Updated Zwicky Catalog of galaxies (Falco et al. 1999) to generate a 
catalog of groups, by means of a friend-of-friend algorithm. The groups
identified have 4 or more members and a surrounding density contrast, 
$\delta\rho/\rho = 80$.
They cover the region $-4\arcdeg<\delta<90\arcdeg$, $|b| >20\arcdeg$ 
spanning over the radial velocity range of 2000 km s$^{-1}$ $<cz<$ 
15000 km s$^{-1}$.   The total sample (hereafter GUZC) comprises 517 groups. 
The correlation length of the total sample is well fitted with a power law
$ \xi(r)=(r/r_0)^\gamma $ with parameters $r_0=9.0 \pm 0.4\ h^{-1}Mpc$  
and $\gamma = -1.67 \pm 0.09$ for values of $r<70\ h^{-1} Mpc$, declining to 
homogeneity at larger scales.  

Three subsamples defined by the range of group virial masses ${\cal M}$ were 
used to have their clustering properties examined throughout the 
autocorrelation function. 
We find an increase of the amplitude of the correlation function according to 
the group masses which extends the results of the $ r_0-d_c$ relation for 
galaxy systems at small $d_c$. 
We obtain $r_0=9.5\pm0.5 h^{-1}Mpc$ ($\gamma=-1.81\pm0.12$), 
$r_0=10.8\pm0.7 h^{-1}Mpc$ ($\gamma=-1.77\pm0.17$), and 
$r_0=14.1\pm1.2 h^{-1}Mpc$ ($\gamma=-1.65\pm0.22$) for mass ranges 
$5\times10^{12}M_{\odot}<{\cal M}<4\times10^{14}M_{\odot} $ ,
$2\times10^{13}M_{\odot}<{\cal M}<4\times10^{14}M_{\odot} $ and
$5\times10^{13}M_{\odot}<{\cal M}<4\times10^{14}M_{\odot} $ respectively.

For completeness we have also analyzed a sample of groups obtained from the
Southern Sky Redshift Survey (da Costa et al.1998) in the range of virial 
masses $5\times10^{12}M_{\odot}<{\cal M}<4\times10^{14}M_{\odot} $ 
to compare the results with  those obtained from GUZC. 
The correlation function of this sample can be fitted with a power law with 
parameters $r_0=8.4 \pm 1.8\ h^{-1}Mpc$ and $\gamma = -2.0 \pm 0.7$.

\end{abstract}


\newpage

\section{Introduction}

The two point correlation function, $\xi(r)$, has been widely used to analyze 
the clustering properties of galaxies and galaxy systems (\cite{peebles93}
and references therein).
Clusters and groups of galaxies can be used as suitable tracers of the 
large scale structure of the universe. In fact, several works have 
characterized the clustering properties of rich clusters of galaxies (see for 
instance \cite{croft97} and \cite{abadi98}) concluding that the cluster 
autocorrelation function has a similar shape than the galaxy correlation 
function with a larger amplitude. 
The clustering properties of groups of galaxies were analyzed by \cite{jing88} 
for the CfA groups (\cite{geller83}), by \cite{maia90} for the SSRS group 
catalog (\cite{maia89}), and by \cite{ramella90} for groups of the CfA slices 
(\cite{huchra90}) where a significant lower amplitude of $\xi(r)$ is detected 
for these systems when compared to rich clusters, as expected in hierarchical 
scenarios of structure formation in the universe. However, the typical values 
found for the correlation length, $r_0$, for the groups examined by these 
authors are significantly different, perhaps due to the small size, and the 
different selection criteria of the samples examined.  In a more detailed work 
\cite{trasarti97} examine how the different criteria of assignment of galaxies 
into  groups affect the determination of $\xi(r)$.  They claim that the 
differences found in previous works may be due to the distinct values of the 
density contrast adopted to generate the catalogs of groups, which resulted in 
groups with distinct physical properties.  

Groups have lower richness and higher space density than the APM, Abell or 
X-ray selected clusters, allowing a detailed analysis of the behavior of the 
$r_0$ at such regimes of space density. It worths to be mentioned that the 
relation between $r_0$ and the mean intergroup separation, $d_c$, for values of 
$d_c$ lesser than $20 h^{-1}Mpc$ has not been extensively explored. 
Moreover, the predictions of current hierarchical models differ from the
universal scaling law proposed by \cite{bahcall92} at large scales 
(\cite{abadi98} and \cite{croft97}), as well as at the smaller scales where 
this work is focused.

In this paper we analyze two new samples of groups of galaxies, the first one, 
taken from the Updated Zwicky Catalog (UZC) by \cite{falco99} which contains 
redshift information for galaxies of the Catalog of Galaxies and Clusters of 
Galaxies (\cite {zwicky61}). The other group catalog, is derived from Southern 
Sky Redshift Survey (SSRS2) by \cite{dacosta98}. Both samples resulted, mainly, 
from the effort of the ON and CfA redshift surveys in the past 20 years, but 
since UZC and SSRS2 come from  distinct galaxy catalogs, we treat both as
independent samples, thus avoiding the mixture of each one biases, also 
allowing us, a cross-check of results.  

Section 2 describes how the groups were defined and which are their general 
properties.  In section 3 we analyze the clustering properties of these groups 
using the autocorrelation function. A brief summary of our conclusions is 
presented in section 4.

\section{Groups in the Updated Zwicky Catalog - GUZC and Southern Sky
Redshift Survey - GSSRS2}

The recent release of the UZC of galaxies (\cite{falco99}), containing 19,369
objects with $m_{Zw} \le 15.5$ and with 96\% of completeness in redshift, 
and the SSRS2 (\cite{dacosta98}), containing 5369 galaxies with $m_B \le 15.5$ 
which covers a region of 1.70 sr of the southern celestial hemisphere.
The large area of sky covered, the relative fainter magnitude limit, 
and the very good completeness in redshift of both catalogs, make them a  
powerful resource to perform statistical analysis of the galaxy properties in 
the nearby Universe.
We used the UZC and SSRS2 to generate new catalogs of groups of galaxies to 
investigate their clustering properties by means of the correlation function. 
The algorithm adopted for the construction of the catalog of groups 
of galaxies is basically the one described by \cite{huchra82}.  Since we are 
interested in "real physical systems", the adopted friends-of-friends algorithm 
(described below), searches for possible group member galaxies, keeping a 
fixed surrounding density contrast ($\delta\rho/\rho$) relative to the mean 
density of galaxies in the catalog. 
 A $\delta\rho/\rho = 80$ was adopted 
because it was demonstrated by \cite{ramella97} to be the best compromise  in 
identifying as many physical loose groups as possible and including all 
systems with high velocity dispersion, but avoiding contamination of the 
catalog by pseudo-groups as well as groups with interlopers (see also 
discussion by \cite{nolthenius87} and \cite{maia90}). 

A search for companions around galaxies is carried out taking into account 
projected separations satisfying 
 
$$D_{12} = 2 \sin\left(\frac{\theta_{12}}{2}\right) \frac{V}{H_0}  \leq D_L$$
 
\noindent and with line-of-sight velocity differences,
 
$$V_{12} = \mid V_1 - V_2 \mid  \leq V_L$$
 
\noindent In the above expressions $V = (V_1 + V_2) / 2$, $V_1$ and $V_2$ are 
the radial velocities of the galaxies, and $\theta_{12}$ their angular 
separation. The quantities $D_L$ and $V_L$ are search parameters scaled 
according to the expressions below in order to take into account the variation 
in the sampling of the galaxy luminosity function, $\phi(M)$, with distance 
 
$$ D_L = D_0 R \quad ; \quad  V_L = V_0 R$$
 
\noindent where
 
$$ R = \left[ \int_{-\infty}^{M_{12}} \Phi (M) dM \bigg/  
  \int_{-\infty}^{M_{lim}}\Phi (M) dM\right]^{-1/3} $$
 
$$ M_{lim} = m_{lim} -25 - 5\log(V_f / H_0) $$
 
$$ M_{12} = m_{lim} -25 -5\log[(V_1 + V_2)/2 H_0] $$

\noindent $D_0$ is the selection parameter at a fixed fiducial radial 
velocity, $V_f$. 
$V_L$ is scaled in the same way as $D_L$.  

To generate the UZC groups (GUZC), the adopted values for $D_0$ and $V_0$ 
were $0.229 h^{-1}Mpc$ and 350 \kms respectively; the apparent magnitude limit, 
$m_{lim} = m_{Zw} = 15.5$, and $H_o = 100h$ \kms Mpc$^{-1}$. The groups present 
a surrounding $\delta\rho/\rho = 80$, for a \cite{schechter76} luminosity 
function adopted for the total sample, parameterized by $\phi^*=0.005$ 
galaxies mag$^{-1}$ Mpc$^{-3}$, $M^*_{B(0)} =-18.97$ and $\alpha=-1.13$ 
(\cite{ramella97}).

The SSRS2 groups (GSSRS2) are obtained with the same $\delta\rho/\rho = 80$, 
but for values for $D_0$ and $V_0$ $0.352 h^{-1}Mpc$ and 350 \kms respectively.
The apparent magnitude, $m_{lim} = m_{B(0)} = 15.5$, while the luminosity 
function  is parameterized by $\phi^*=0.0137$ galaxies mag$^{-1}$ Mpc$^{-3}$,  
$M^*_{B(0)} =-19.40$ and $\alpha=-1.08$ (\cite{marzke98}).

We have also computed the virial masses of the groups combining the virial 
radius and the velocity dispersion ${\cal M}\propto R_v \sigma^2$ which provide
suitable estimates of group masses (\cite{ramella97}).
According to the same authors, several groups with 3 or 4 members may be 
unreliable, although we should expect a low degree of contamination 
by pseudo-groups (as small as 10\%) by adopting $N\ge4$.

The resulting group catalogs contain systems with at least 4 members and 
mean radial velocities, $V_g \le$ 15,000 \kms (GUZC), and $V_g \le$ 12,000 \kms 
(GSSRS2) consisting in the largest sample of groups of galaxies homogeneously 
selected to the present.
For analysis purposes, we removed groups in regions of 
galactic latitude  $|b| < 20^o$ and declination $\delta > -4^o$ for GUZC and 
for the GSSRS2 we only considered groups within the following limits: 
$-40^o < \delta < -2.5^o$ and $b < 40^o$  (southern galactic cap) and 
$\delta < 0^o$ and $b < +35^o$ (northern galactic cap).
This choice provides a sample free of strong obscuration from the Galaxy and 
also avoids possible lack of homogeneity in the Falco catalog due to the not 
complete coverage of the Zwicky catalog in part of those regions.
For both them, very rich groups
containing more than 40 galaxies were also removed, to avoid the presence of 
rich clusters in our analysis.  
In addition to the above criteria, those groups with $V_g \le$ 2,000 \kms were
also discarded from the analysis, to prevent any significant contribution of 
the group peculiar velocities in the measured redshifts which were used to 
determine their distances.  

The projected distribution of the GUZC (panel a) and GSSRS2 (panel b) groups in 
equatorial coordinates is shown in figure 1.  
Histograms of the distributions of $V_g$ and group mass (${\cal M}$) for 
the systems of the GUZC (panel a) and GSSRS2 (panel b) are presented in 
figures 2 and 3 respectively.  
The velocity distribution of groups is similar to the one for galaxies, and a more 
detailed examination of the estimated masses for groups between 2000 \kms and 
5000 \kms confirms the reliability of this determination, in the sense that groups 
with lower masses are made up of few bright galaxies or, if they have a large number 
of members, they also contain a higher fraction of intrinsically less luminous 
objects.

\section{Analysis}

The lower richness and higher space densities of GUZC and GSSRS2 groups, compared to 
the APM or X-ray selected clusters allow the evaluation of the dependence of 
$r_0$ with $d_c$ to be examined for higher regimes of spatial densities in order to 
extend the available data of this relation and to compare with the $r_0 = 0.4\ d_c$ 
law (\cite{bahcall92}).  

The spatial correlation function is computed using the estimator given by 
\cite{peebles93}

$$
\xi(r)=\frac{DD(r)}{DR(r)}\frac{N_R}{N_D}-1
$$

\noindent where $DD(r)$ and $DR(r)$ are the number of real and random pairs of 
groups at separation $r$ and $N_D$ and $N_R$ are the number of objects in the 
data and random catalogs, respectively. We use an equal weighting of
each data/random point for simplicity given that error bars in our samples 
are significantly larger than the expected differences between the estimates 
of $\xi(r)$ from different weighting schemes according to the analysis by 
\cite{ratcliffe98}.  
The catalogs of random groups where generated within the same volume as the 
real groups, with a velocity distribution corresponding to the best 
third-degree polynomial fit to each $V_g$ distribution.  
The choice on the actual group distribution rather than that for the
galaxies, is motivated by the fact that even though both present similar 
behavior, by using the group distribution it is possible to avoid biases 
caused by the distribution of galaxies in low density regions devoid of richer 
groups.  
 Errors were derived from the bootstrap resampling technique developed by 
\cite{barrow84} with 30 bootstrap samples. 

We estimated $\xi(r)$ for three samples, taking into account different intervals of 
group masses, ${\cal M}$, which represents, also, three distinct $d_c$ values.  
In order to deal with reasonable number of systems in the subsamples, the adopted 
mass limits are: 
$5\times10^{12}M_{\odot}<{\cal M}<4\times10^{14}M_{\odot} $,
(397 groups, sample 1); 
$2\times10^{13}M_{\odot}<{\cal M}<4\times10^{14}M_{\odot} $,
(268 groups, sample 2); and 
$5\times10^{13}M_{\odot}<{\cal M}<4\times10^{14}M_{\odot} $,
(167 groups, sample 3). 
The lower limit in the first sample corresponds to the typical mass of the group with 
four $L_*$ members, the upper limit in the three samples was chosen to avoid groups 
with masses greater than the estimated mass of the Virgo cluster, (\cite{ramella97}).

The estimated $\xi(r)$ for the total sample of GUZC groups is shown if figure 4.   
The correlation function gives a positive signal up to $70 h^{-1}Mpc$ and it has a 
similar shape to the one for the galaxies.   The results for the three constrained 
subsamples are shown in figure 5, 6 and 7 for the range of masses corresponding to 
samples 1, 2 and 3 respectively.
By the inspection of figures 4, 5, 6 and 7 it is possible to notice that the behavior 
of $\xi(r)$ for all the four samples is consistent with a power-law.  The amplitude 
of the correlation function presents a tendency to increase, as far as we have a 
predominance of more massive groups. 
We have used the Levenbergh-Marquardt method, which takes into account errors with 
minimum non-linear least-squares, to estimate the best fitting parameters 
$r_0$ and $\gamma$ in the power-law approximation $(r/r_0)^\gamma$.   We obtain 
$r_0=9.5\pm0.5 h^{-1}Mpc$ ($\gamma=-1.81\pm0.12$), $r_0=10.8\pm0.7 h^{-1}Mpc$ 
($\gamma=-1.77\pm0.17$), and $r_0=14.1\pm1.2 h^{-1}Mpc$ ($\gamma=-1.65\pm0.22$) 
for subsamples 1, 2 and 3 respectively. 

In order to test the stability of the estimates of the correlation function
against the existence of possible fake groups, we have removed from the
total sample those groups with four members and mass exceeding 
$5\times10^{13}M_\odot$ (48 groups). These massive groups with few members 
are subject to suspicion and could strongly contribute to the expected 10\% 
fraction of pseudo-groups as discussed by \cite{ramella97}. The correlation 
function for this restricted sample ($r_0=(9.0\pm0.4)h^{-1}Mpc, 
\gamma=1.72\pm0.10$) is 
indistinguishable within errors from that of the total sample, indicating 
that fake groups are not likely to generate strong systematic effects in our 
sample.

Using the same procedure we have computed $\xi(r)$ for the 104 groups in 
the GSSRS2 with the results displayed in figure 8.  The best fitting 
parameters are $r_0=8.4\pm1.8$h$^{-1} Mpc$ and $\gamma=-2.0\pm0.70$.  
Due to the small number of objects in this sample we have not attempted 
divisions into subsamples. 

A comparison of our results with those obtained by \cite{jing88}, 
\cite{maia90}, \cite{ramella90}, and \cite{trasarti97} is presented in 
Table 1.  The $r_0$ and $\gamma$ values they have found, are different 
when compared to the one we have found for the total sample.  
The possible causes of the differences may be related to the small number 
of objects in the previous samples, to the minimum number of member galaxies 
assigned to groups, and also to the different $\delta\rho/\rho$ adopted for 
the definition of groups, which results in the determination of groups with 
distinct physical properties.  
The different characteristic depths of the samples may also 
play an important role in the range of fitting $\xi(r)$.  
In particular the samples analyzed by \cite{jing88} and by \cite{maia90}, 
which retain most of those characteristics described above, are the ones 
which present the most discrepant results.

We have estimated the values of $d_c$ for the subsamples in two different ways:
(a) we compute the abundance $n(>{\cal M})$ of systems using \cite{bahcall93} 
mass function, in the corresponding range of masses and estimate 
$d_c=n^{(-1/3)}$. (b) we consider that the GUZC and GSSRS2 catalogs are 
bona-fide complete out to 5000 \kms and we fit a homogeneous distribution 
in the range  2000 \kms $< V_g <$ 5000 \kms.
Taken into account \cite{ramella97}  we expect the loss of groups at
5000 km/s with at least 4 members less than 20\%, so the second method
should also provides a reliable estimate of $d_c$.

The resulting values for GUZC are: $d_c=[12]\ h^{-1}$ Mpc for the total
sample, while $d_c=9\ [14]\ h^{-1}$ Mpc, $d_c=15\ [16]\ h^{-1}$ Mpc and 
$d_c=21\ [22]\ h{-1}$ Mpc 
for subsamples 1, 2 and 3 respectively.  
For the GSSRS2, $d_c=9\ [12]\ h^{-1}$ Mpc in the range of masses 
$5\times10^{12}M_{\odot}<{\cal M}<4\times10^{14}M_{\odot} $. 
The numbers in square brackets correspond to the estimates using method (b).
Both methods give consistent values of $d_c$, provided that the typical 
uncertainties in $d_c$ are $\simeq 2 h^{-1}$ Mpc.

In figure 9 we plot the derived values of $r_0$ as a function of $d_c$ for the 
different samples of groups considered. Also, are plotted in this figure the 
results for APM clusters according to \cite{croft97}.  
All the group sets present $r_0$ values higher than the universal scaling 
law $r_0=0.4 d_c$ by \cite{bahcall92} in contrast to the clusters  which have a 
lower amplitude than predicted by this law.
Meanwhile, it is noticeable in this figure the shape of the $r_0-d_c$ 
relation is similar to that expected in hierarchical models for structure 
formation such as a cold dark matter scenario discussed by \cite{croft97}. 

\vskip 1cm

\section{Conclusions}

We have analyzed the two point correlation function for two new catalogs of 
groups of galaxies. The GUZC catalog is the largest homogeneous sample to the 
present and provides a suitable data set to analyze the clustering properties 
of systems of galaxies of low richness.
The group correlation function, for the total sample of groups, is well fitted 
to a power-law of the form $\xi=(r/r_0)^\gamma$, with $\gamma=-1.67\pm 0.09$ 
and $r_0=9.0 \pm 0.4$ h$^{-1}$ Mpc for values of $r< 70 h^{-1}$ Mpc.
A similar analysis for GSSRS2 catalog produces $ r_0=8.4\pm1.8$h$^{-1}$ Mpc
and $\gamma=-2.0\pm0.7$.
These results differ from 
the findings of other authors mainly due to the different groups 
characteristics of the samples analyzed. 

The results for subsamples of groups with different group mass intervals, 
show the tendency for higher values of $r_0$ at the small mean intergroup 
separations examined, when compared with the universal scaling law proposed 
by \cite{bahcall92}. 

Correlation functions of APM clusters (\cite{croft97}), X-ray confirmed Abell
clusters (XBACs, \cite{abadi98}) and the results for groups presented in this 
paper strongly argue for a hierarchical scenario of structure formation in the 
universe such as cold dark matter models.

This paper was partially supported by CONICET, SeCyT UNC and Fundaci\'on 
Antorchas, Argentina.  MAGM acknowledges CNPq grant 301366/86-1.

\vskip 0.5cm

\centerline{ NOTE ADDED IN PROOF}

After this paper was submitted to ApJ, \cite{girardi}
examined $\xi(r)$ of groups relative to the one for 
galaxies using the UZC and SSRS2 catalogs of galaxies. They find similar 
results to ours for a volume-limited sample of 139 groups 
($r_0=(8 \pm 1)h^{-1}$Mpc; $\gamma=-1.9\pm0.7$).

 {}

\newpage
\begin{deluxetable}{lccccc}
\small
\tablecolumns{6}
\tablenum{1}
\tablecaption{Results of clustering analysis for several catalogs of groups.}
\tablehead{
\colhead{ Sample } &
\colhead{ $r_0\ [h^{-1}Mpc]$ } &
\colhead{ $\gamma$ } &
\colhead{ $d_c\ [h^{-1}Mpc]$ } &
\colhead{ \# of groups } 
}
\startdata
Total Sample &  $9.0\pm0.4$   & $-1.67 \pm 0.09$ & $ - $ [$12$] & $517$ \nl

Sample 1     &  $9.5\pm0.5$  & $-1.81 \pm 0.12$ & $ 9 $ [$14$] & $397$ \nl

Sample 2     &  $10.8\pm 0.7$ & $-1.77 \pm 0.17$ & $15 $ [$16$] & $268$  \nl

Sample 3     &  $14.1\pm 1.2$ & $-1.65 \pm 0.22$ & $21 $ [$22$] & $167$ \nl

GSSRS2       &  $8.4 \pm 1.8$ & $-2.0  \pm 0.7$  & $ 9 $ [$12$] & $104$ \nl

\cite{jing88} & $3.5        $ & $-1.77        $  & $-  $        & $174$ \nl

\cite{maia90} & $3.2        $ & $-1.82        $  & $9.1$        & $87$  \nl

\cite{ramella90}& $6.0      $ & $-1.0         $  & $- $         & $128$ \nl

\cite{trasarti97}& $ 7.03 \pm 0.6$ & $-1.21 \pm 0.04$ & $- $ & $\approx 200$ \nl
\enddata
\label{table1}
\tablenotetext{}{Quoted values of $d_c$ were calculated using \cite{bahcall93} 
mass function, numbers in brackets correspond to the assumption of a 
homogeneous distribution in the range $ 2000$ \kms $\le V_g \le 5000$ \kms.} 
\end{deluxetable}

\newpage

\begin{figure}[ht] 
\centerline{\psfig{file=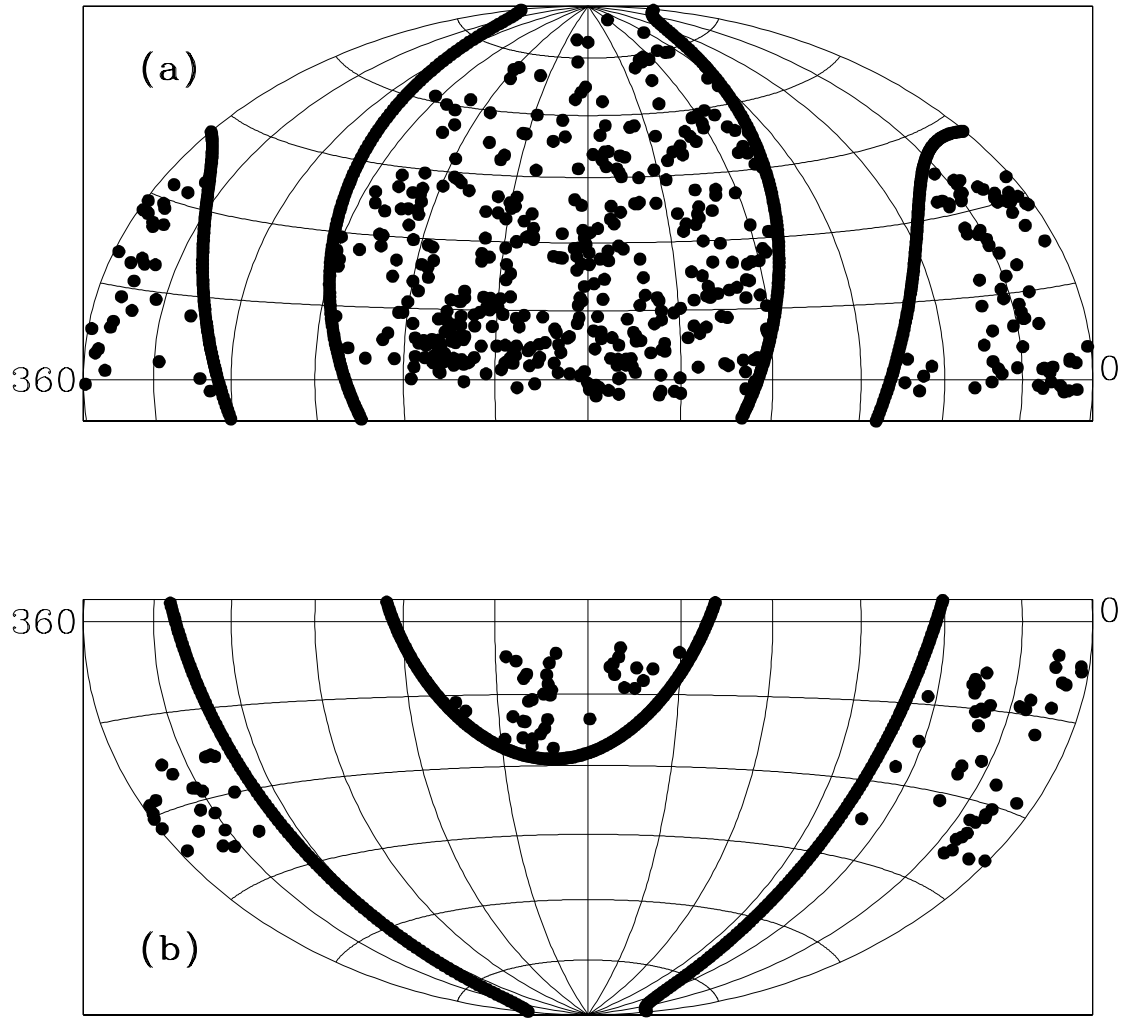,width=16cm}}
\caption{Projected distribution of GUZC (panel a) and GSSRS2 (panel b) 
groups.  Solid lines correspond to the limit in galactic latitude $|b|=20^o$ 
(GUZC) and $|b|=35^o$ (limit of SSRS2 catalog) used in the statistical 
analysis.}
\label{fig1} 
\end{figure}

\begin{figure}[ht] 
\centerline{\psfig{file=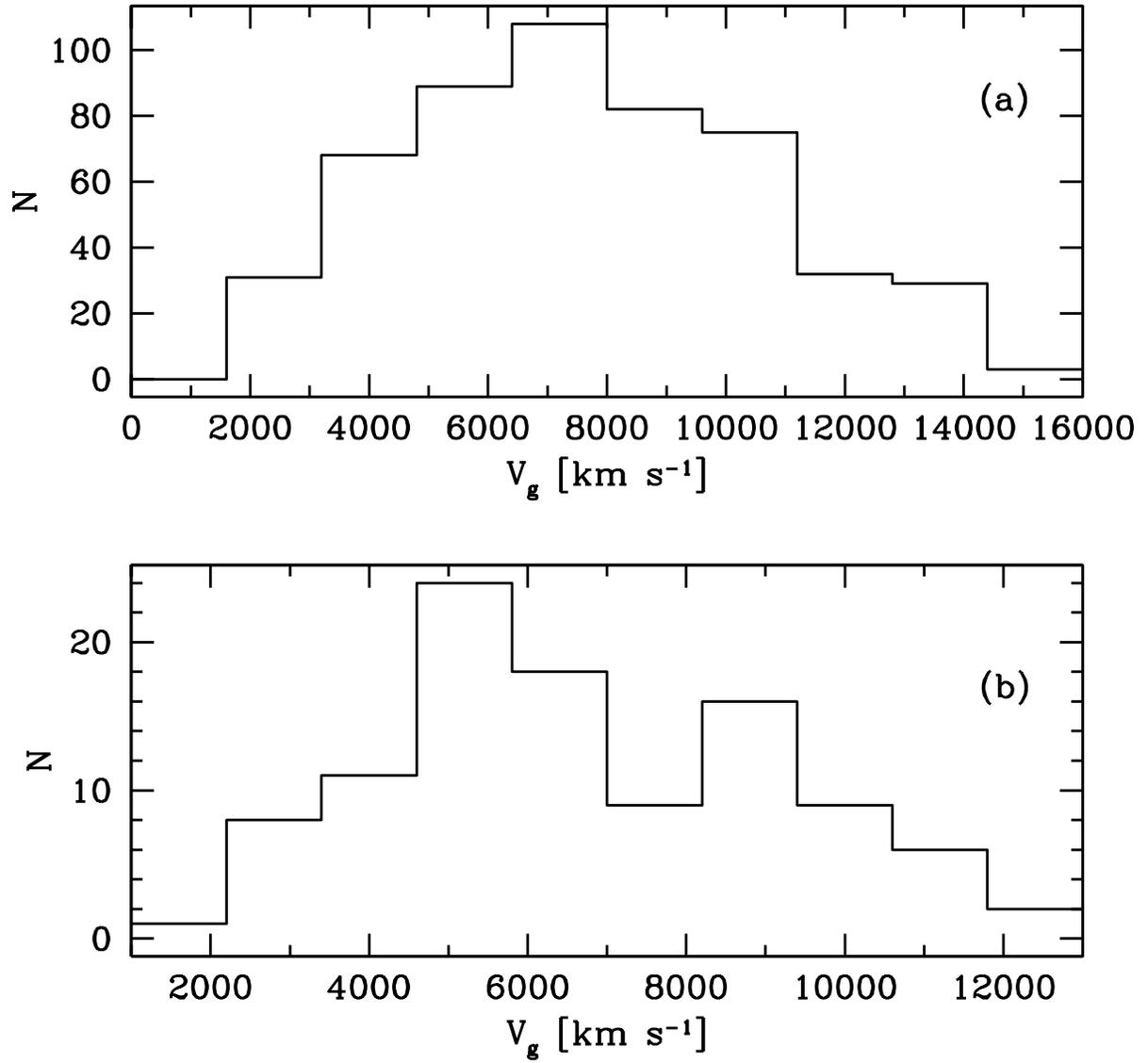,width=16cm}}
\caption{Distribution of mean radial velocities ($V_g$) of GUZC (panel a) 
and GSSRS2 (panel b) groups.}
\label{fig2} 
\end{figure}

\begin{figure}[ht] 
\centerline{\psfig{file=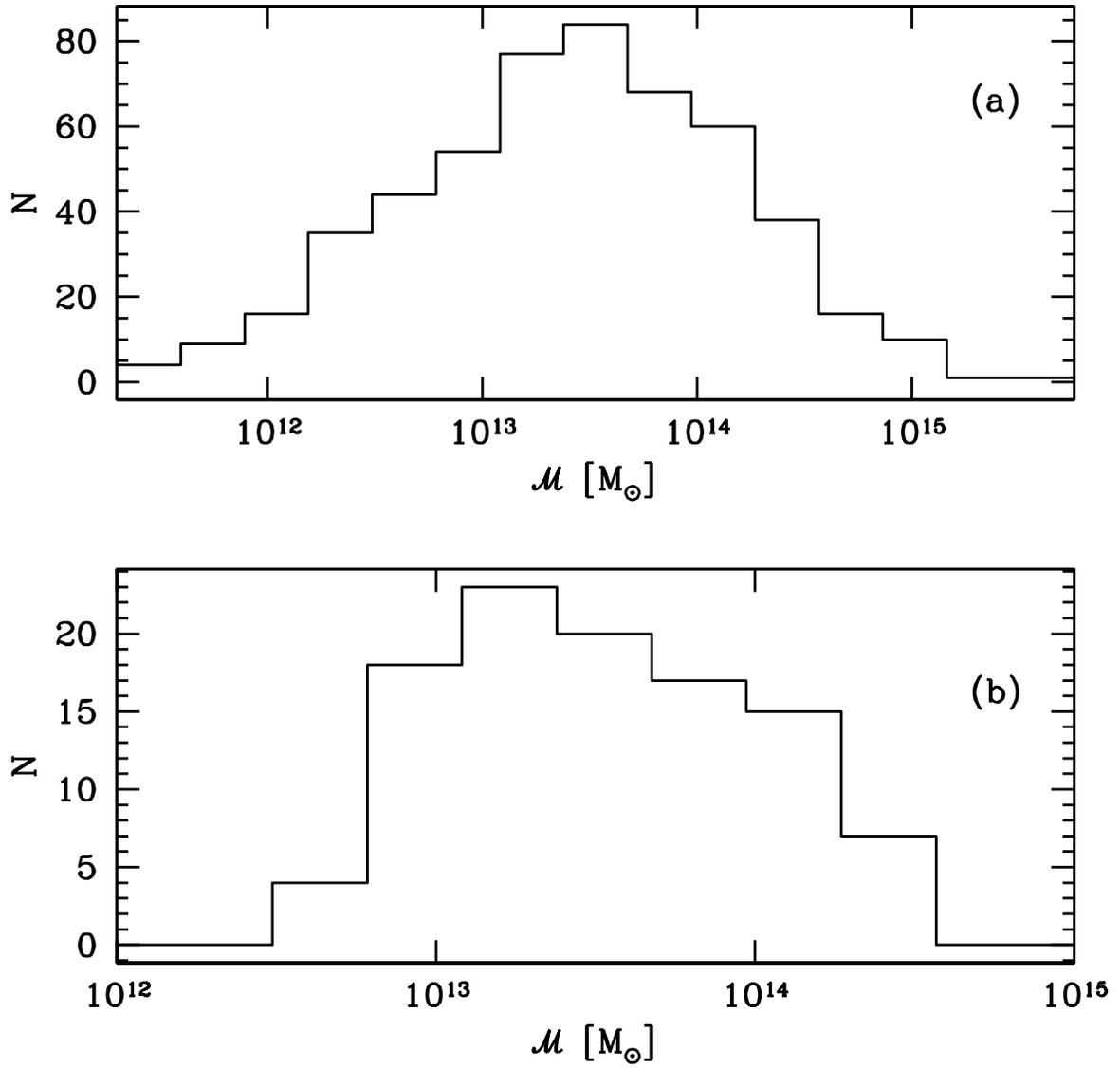,width=16cm}}
\caption{Distribution of virial masses (${\cal M}$) for GUZC (panel a) and
GSSRS2 (panel b) groups.}
\label{fig3} 
\end{figure}

\begin{figure}[ht] 
\centerline{\psfig{file=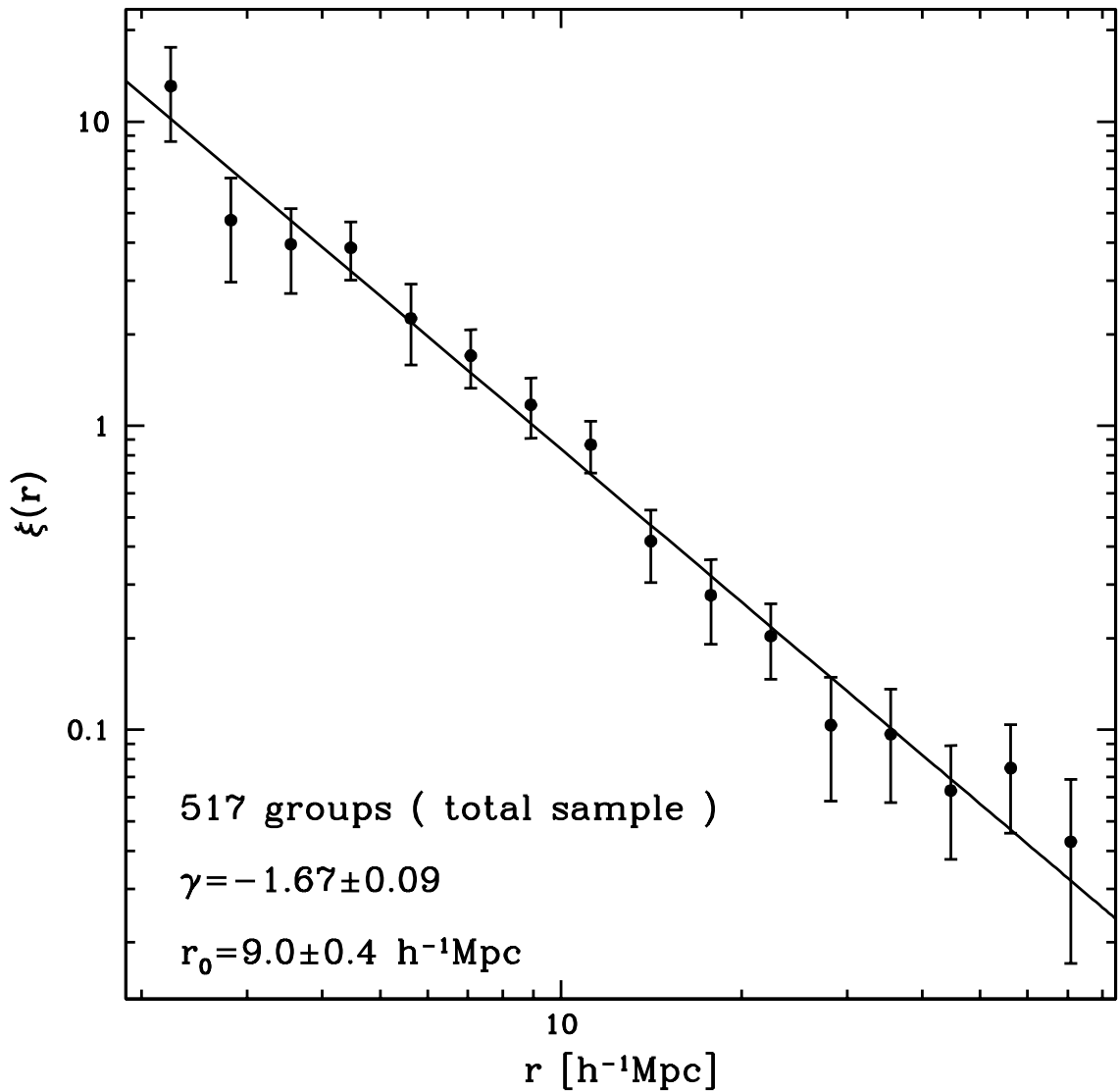,width=16cm}}
\caption{Spatial autocorrelation function, $\xi(r)$, for the total GUZC 
groups (517 objects). Error bars correspond to bootstrap resampling estimates 
of uncertainties. The solid line shows the power law fit $(r/r_0)^\gamma$ 
with $r_0=9.0\pm 0.4$\ h$^{-1}$Mpc and $\gamma = -1.67 \pm 0.09$.} 
\label{fig4} 
\end{figure}

\begin{figure}[ht] 
\centerline{\psfig{file=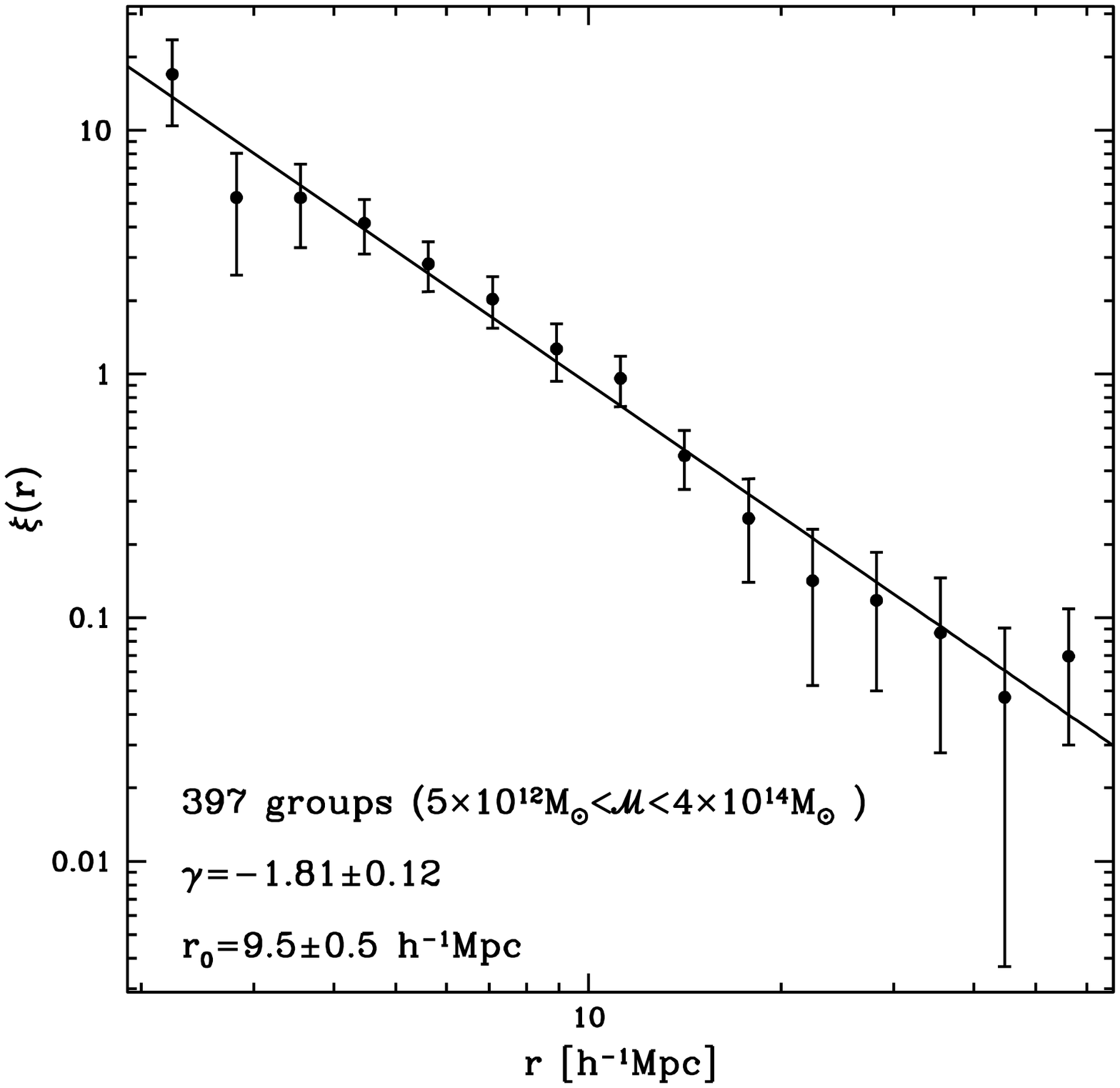,width=16cm}}
\caption{The same as figure 4 for GUZC groups with 
$5 \times 10^{12} < {\cal M} < 4 \times 10^{14}$.}
\label{fig5} 
\end{figure}

\begin{figure}[ht] 
\centerline{\psfig{file=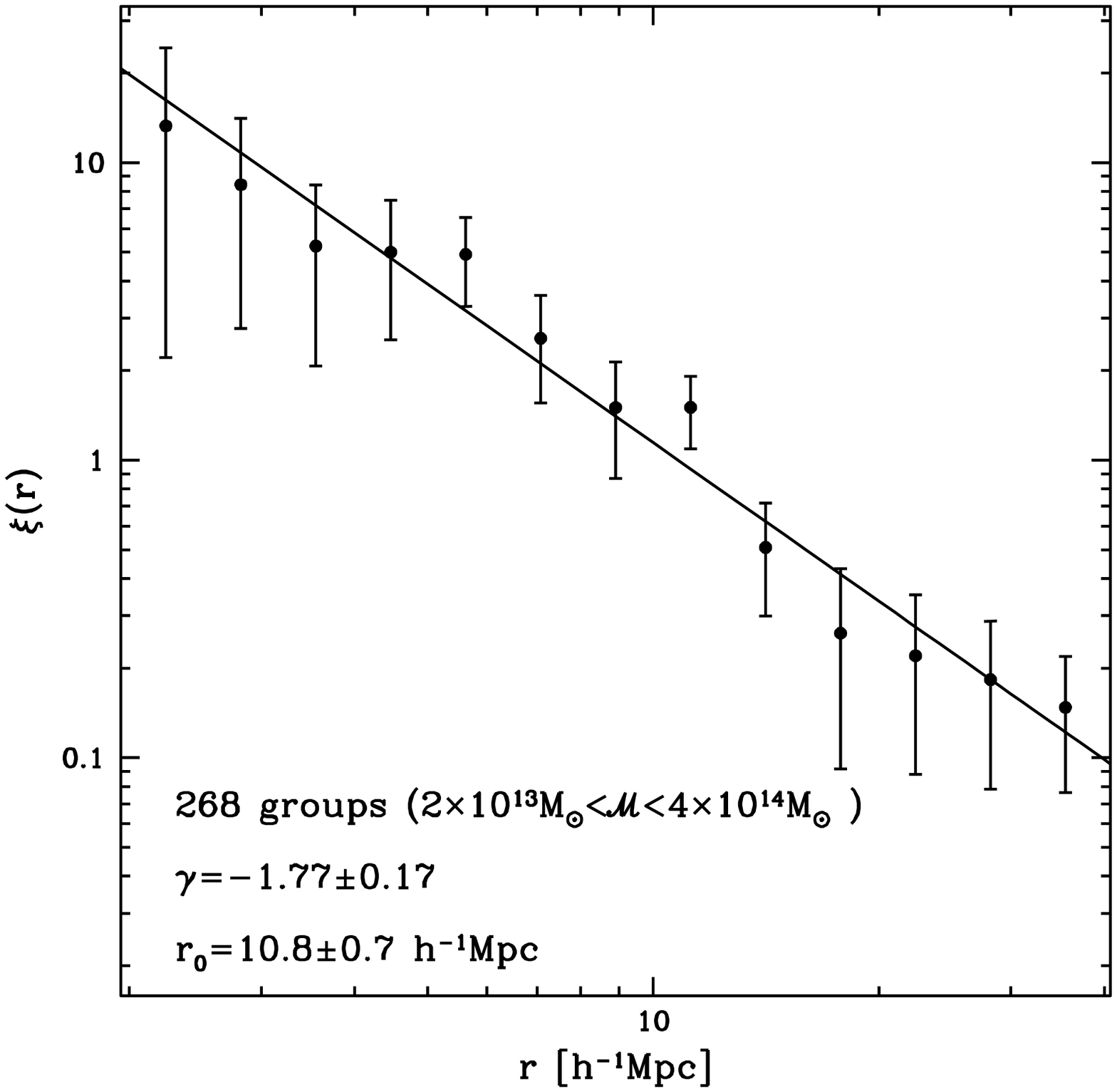,width=16cm}}
\caption{The same as figure 4 for GUZC groups with 
$2 \times 10^{13} < {\cal M} < 4 \times 10^{14}$.}
\label{fig6} 
\end{figure}

\begin{figure}[ht] 
\centerline{\psfig{file=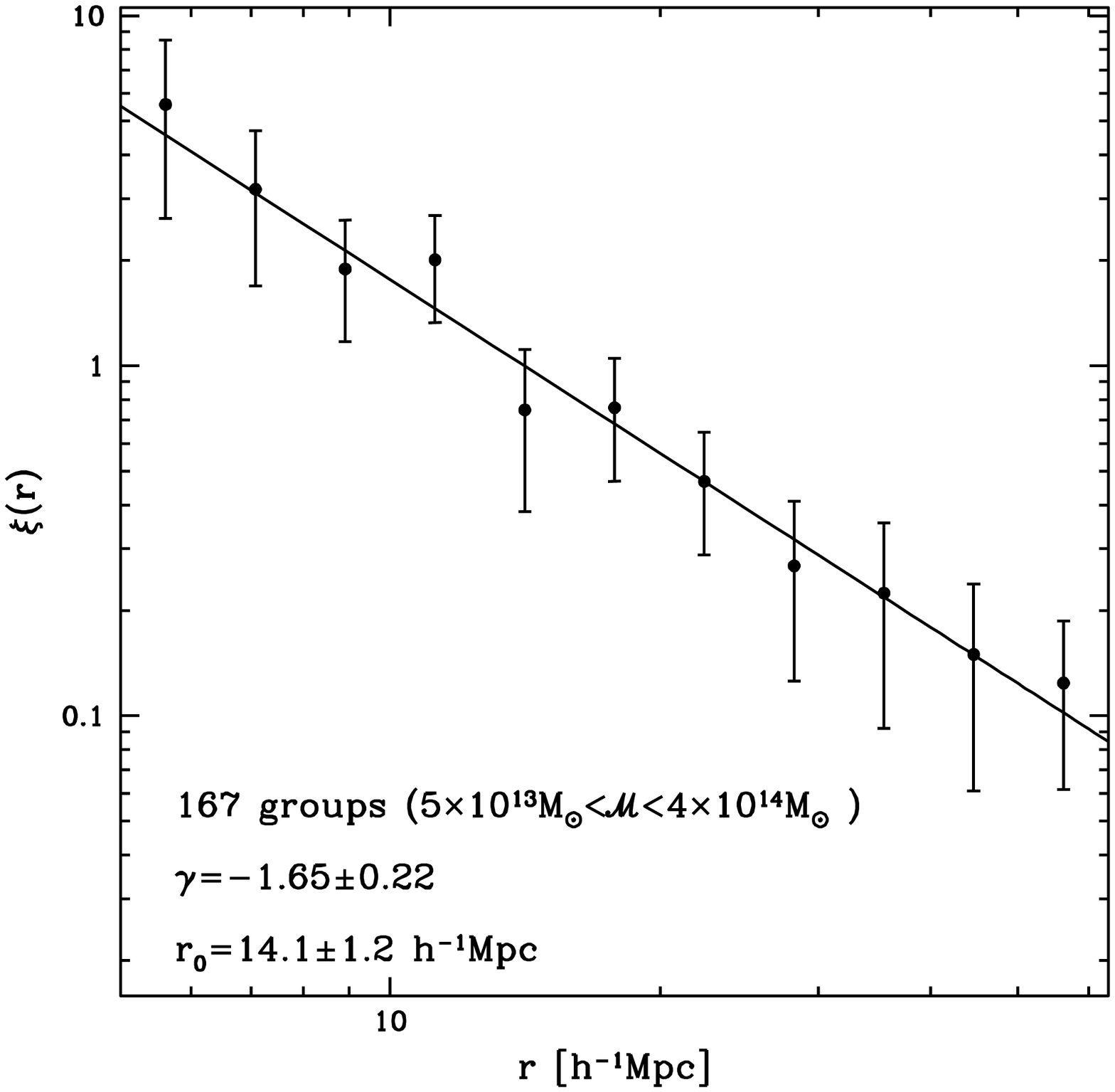,width=16cm}}
\caption{The same as figure 4 for GUZC groups with 
$5 \times 10^{13} < {\cal M} < 4 \times 10^{14}$.}
\label{fig7} 
\end{figure}

\begin{figure}[ht] 
\centerline{\psfig{file=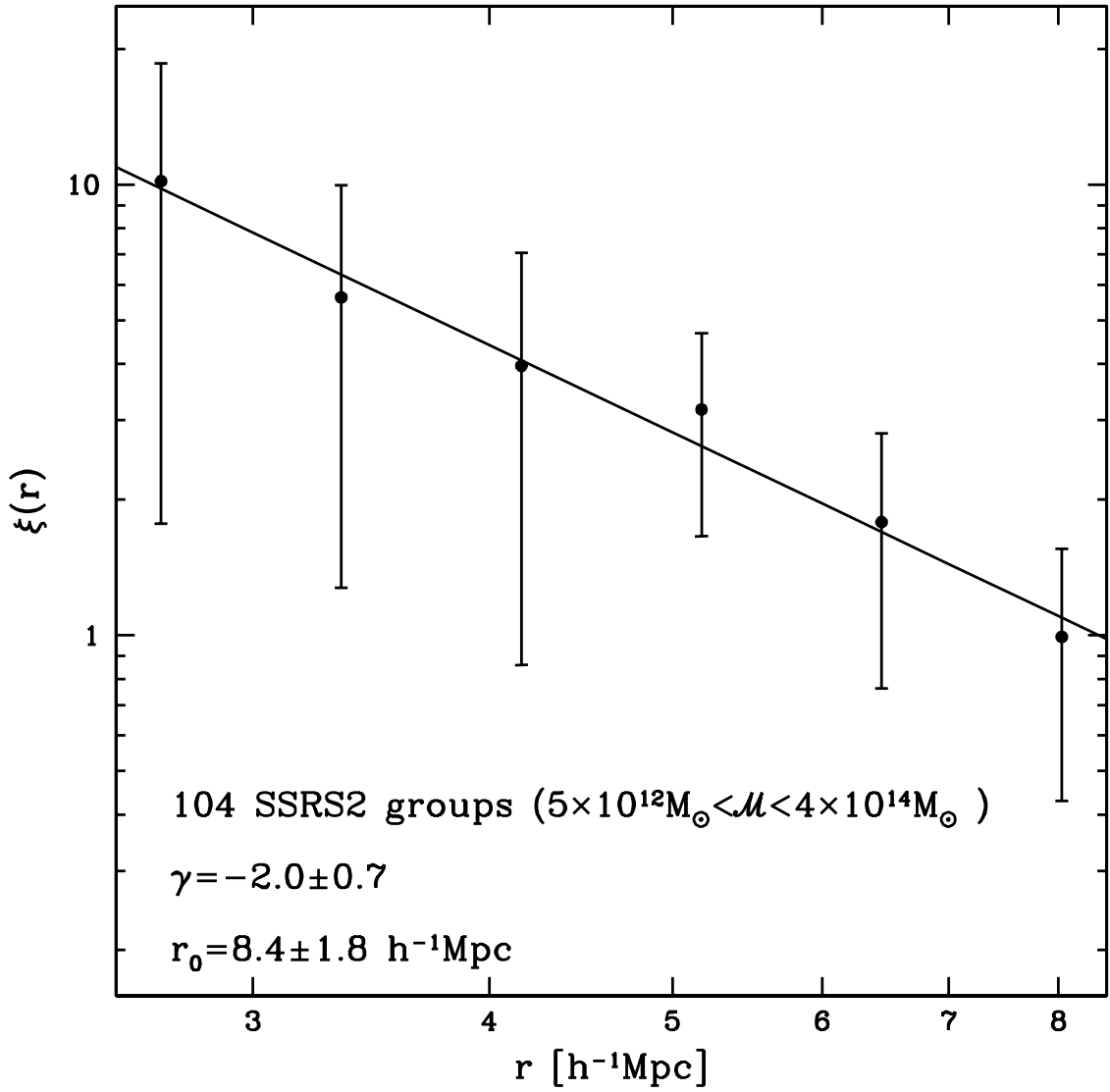,width=16cm}}
\caption{The same as figure 4 for GSSRS2 groups with 
$5 \times 10^{12} < {\cal M} < 4 \times 10^{14}$.}
\label{fig8} 
\end{figure}

\begin{figure}[ht] 
\centerline{\psfig{file=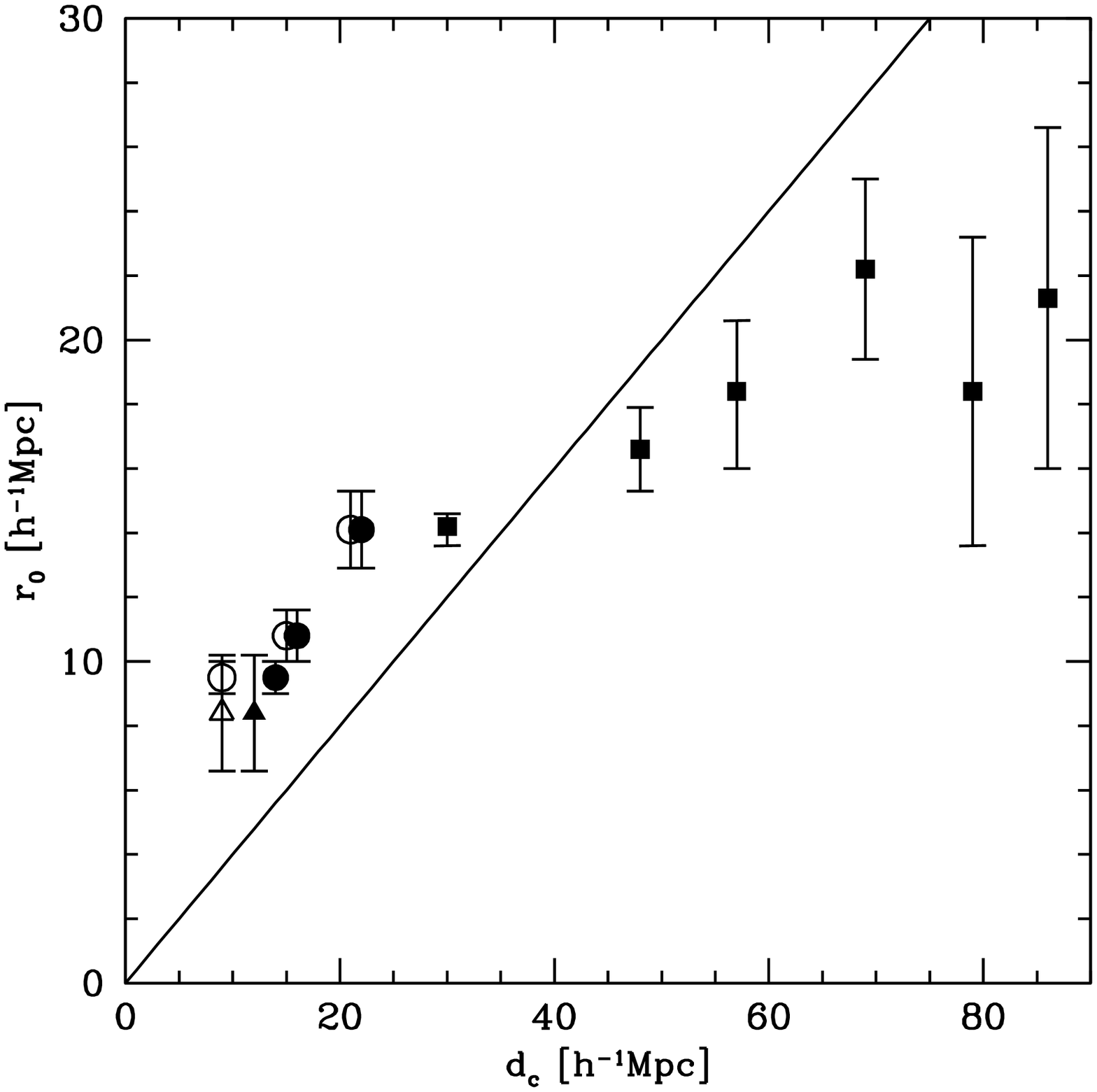,width=16cm}}
\caption{Correlation length, $r_0$, as a function of the mean intergroup 
separation, $d_c$. Circles correspond to the three GUZC subsamples of groups 
analyzed, while triangles represents the values for the GSSRS2 sample. Open 
symbols correspond to $d_c$ estimates from \cite{bahcall93} mass function, 
filled symbols to our estimates assuming completeness in 
$ 2000$ \kms $\le V_g \le 5000$ \kms.   
Squares show the $r_0-d_c$ relation of APM clusters as described by 
Croft et al. 1997, while the line represents the universal scaling law 
$r_0 = 0.4 d_c$. The error bars for our points in the figure correspond to 
those derived from the non linear least-squares fit.} 
\label{fig9} 
\end{figure}

\end{document}